\documentclass{aa}  

\usepackage{graphicx}
\usepackage{lmodern}
\usepackage{txfonts}
\usepackage{natbib}
\usepackage[colorlinks=true,linkcolor=blue,citecolor=blue,urlcolor=blue]{hyperref}
\usepackage[capitalize]{cleveref}
\usepackage{xcolor} 
\begin{document} 

   \title{Weighing Galaxies Inside-Out: Small-Scale Lensing and the Stellar Mass Problem}

\titlerunning{Studying stellar content of the galaxies using small-scale weak gravitational lensing}
\author{Divya Rana\inst{1}\and
        Henk Hoekstra\inst{1}\and
        Konrad Kuijken\inst{1}\and
        Alessandro Sonnenfeld\inst{2}
        \and Maryam Tajalli\inst{3}}%

\institute{
Leiden Observatory, Leiden University, P.O. Box 9513, 2300 RA Leiden, The Netherlands
\email{rana@strw.leidenuniv.nl, hoekstra@strw.leidenuniv.nl}
\and
Department of Astronomy, School of Physics and Astronomy, Shanghai Jiao Tong University, Shanghai 200240, China;
State Key Laboratory of Dark Matter Physics, School of Physics and Astronomy, Shanghai Jiao Tong University, Shanghai 200240, China;
Shanghai Key Laboratory for Particle Physics and Cosmology, Shanghai 200240, China
\and
Max Planck Institute for Astrophysics, Karl-Schwarzschild-Stra\ss{}e 1, 85748 Garching bei M\"unchen, Germany
}

\date{\today}





\abstract
{Small-scale galaxy--galaxy weak lensing provides an independent way to probe the stellar initial mass function by constraining the matter distribution within galaxies. We forecast the potential of upcoming \textit{Euclid} observations using realistic mock catalogues designed for a DESI-like Bright Galaxy Survey, which is expected to cover approximately $9000\,\mathrm{deg}^2$ of the \textit{Euclid} wide survey area. We assess the impact of foreground lens light, which introduces significant scale-dependent biases in source detection, photometry, and shape measurements. Subtracting the lens light with \textsc{Galfit} reduces these biases by nearly an order of magnitude at scales down to four times the half-light radius. We model the predicted lensing signal around central galaxies with stellar masses between $9.5 \leq \log(M_*/h^{-2}M_\odot) \leq 11.5$ over projected separations of $10$ to $100\,h^{-1}\mathrm{kpc}$. The resulting signal-to-noise ratios reach $\sim 150$ for intermediate-mass galaxies. Assuming an NFW profile for the dark matter contribution, we constrain the stellar mismatch parameter $\alpha$ to a precision of approximately $8.5$ percent for the full sample and $8.6$ percent for massive galaxies. These results highlight the potential of small-scale galaxy--galaxy weak lensing as a robust probe of the IMF and the connection between stellar and dark matter in galaxies using upcoming \textit{Euclid} data.}

 

   \maketitle
%
\section{Introduction}

The stellar initial mass function (IMF) describes the mass distribution of stars at their formation. It serves as a crucial basis for converting observed galaxy photometry and spectroscopy into physical stellar masses through stellar population synthesis (SPS) modeling \citep[e.g.,][]{1968_Tinsley,2009_Conroy,2013_Conroy}. Traditionally regarded as a universal parameter, recent observational studies increasingly suggests that the IMF varies depending on the local galactic environment \citep{2010_Bastian,2010_Treu,2012_Cappellari,2013_Geha} and may also evolve over different cosmic epochs \citep{2012_Conroy,2018_Zhang}.

These environmental factors are linked to broader properties of galaxies, such as metallicity, mass-to-light ratio, and stellar velocity dispersion \citep{2004_Portinari,2016_Clauwens,2021_Goswami}. 
Notably, massive early-type galaxies (ETGs) show evidence for a non-universal initial mass function (IMF) \citep[for a review, see][]{2020_smith}. Below $\sim1\,M_\odot$, their spectra are best reproduced by an IMF steeper than that of the Milky Way (i.e.\ bottom-heavy, with a higher proportion of low-mass stars). This signal, however, may be confined to the innermost $\lesssim2$\,kpc, and some strong-lensing measurements remain consistent with a Milky Way-like IMF \citep{2015_Smith_russell}. Above $\sim1\,M_\odot$, by contrast, a slightly shallower (top-heavy) slope is preferred. This is in line with the top-heavy IMFs inferred in intense starbursts, the environments in which the cores of massive ETGs are thought to have formed \citep[e.g.][]{2018_Zhang}. As a result, the idea of a universally applicable IMF—a notion that is still present in many theories of galaxy evolution—is increasingly being challenged \citep[e.g.,][]{2010_Van_Dokkum,2012_Spiniello,2013_Ferreras}.

The effects of a variable IMF are especially pronounced at high redshift. Galaxies from this early era formed under conditions characterized by lower metallicities, higher gas densities, and increased star formation rates, which contrast sharply with those in the local Universe \citep{1998_Larson, 2005_Weidner, 2012_Marks, 2014_Chabrier, 2018_Hopkins}. Thus, using local IMF templates for distant galaxies can lead to significant systematic errors in the calculation of stellar masses, star formation histories, and chemical evolution profiles \citep{2010_Bastian}.

The impact of systematic biases significantly affects cosmology and galaxy formation theory. The stellar-to-halo mass relation, which connects baryonic galaxies to their corresponding dark matter haloes, relies heavily on accurate stellar mass estimates \citep{2019_Posti}. Modern cosmological hydrodynamical simulations calibrate intricate sub-grid baryonic processes—such as active galactic nucleus (AGN) activity and supernova feedback—to replicate large-scale observables like the galaxy stellar mass function \citep[GSMF; e.g.,][]{2023_Schaye,2023_Kugel,2026_Schaye}. A variable IMF can fundamentally change the estimated mass-to-light ratios of individual galaxies, thereby distorting both the shape and normalization of the derived GSMF. If these simulations are calibrated using a GSMF that assumes a universal IMF, the resulting observational biases will directly influence the feedback efficiencies they tune, leading to systematics in their predictions regarding baryonic assembly and cosmological observables.

Independent investigations into the total mass distribution are essential. Dark matter-only $\Lambda$ cold dark matter ($\Lambda$CDM) simulations indicate that density profiles become cuspy towards the centre, scaling as $\rho(r) \propto r^{-\gamma}$, with $\gamma \lesssim 1$ at the innermost resolved radii \citep{2010_Navarro, 2025_Enomoto}. In contrast, kinematic observations of dwarf and low surface brightness galaxies frequently show much shallower density cores \citep[e.g.][]{2001_Mcgaugh,2005_Simon,2008_Kuzio}, whereas strong lensing combined with kinematics finds inner dark matter slopes consistent with the Navarro--Frenk--White \citep[NFW;][]{1997_Navarro} profile in massive elliptical galaxies at $z \approx 0.35$ \citep{2025_Sheu}. Determining whether this discrepancy between core and cusp density profiles is due to violent baryonic feedback \citep{2012_Maccio} or suggests modifications to the fundamental nature of dark matter, such as warm or self-interacting dark matter \citep{2000_Spergel,2016_Bose,2018_Tulin}, requires robust observational constraints that are independent of stellar population synthesis (SPS) models.

In a galaxy, the gravitational potential shifts from being dominated by baryonic matter in the dense core to dark matter in the extended halo, with this critical transition occurring around the effective radius. Studying this transition is vital for our understanding, but even slight changes to the chosen IMF can lead to variations in the inferred stellar mass by as much as a factor of two, despite analysing the same data \citep{2013_Barnab,2014_Courteau,2017_Wright}. Addressing these uncertainties is therefore essential for advancing our knowledge of galactic dynamics and structure.

The total mass of galaxies can be studied observationally. For instance, stellar kinematics can provide dynamical mass profiles \citep{2012_Cappellari}, while strong gravitational lensing accurately measures the mass within the Einstein radius \citep[e.g.,][]{2010a_Treu,2024_Shajib}. By combining methods,
precise  constraints on kiloparsec scales have been obtained   \citep[e.g.,][] {2007_Gavazzi,2007_Johnston,2013_Newman,2014_Oguri,2019_Sonnenfeld}. However, strong lensing events are intrinsically rare and tend to be biased towards massive ellipticals at intermediate redshifts, which limits their effectiveness for comprehensive, population-wide studies, while dynamical studies need to make assumptions about the orbits of stars.

Galaxy-galaxy weak lensing effectively overcomes these limitations by statistically characterizing the mass distribution across representative samples of galaxies. The gravitational potential of foreground galaxies distorts the shapes of background sources. By stacking this subtle shear signal around large samples of foreground galaxies, the averaged total mass profile can be accurately mapped, without prior assumptions about the IMF. 
Therefore, weak lensing has become a fundamental technique for studying halo structures, the connection between galaxies and their host halos, and for precision cosmology \citep[e.g.,][]{Mandelbaum2006,Leauthaud2012,Hudson2015,Sonnenfeld2019,Dvornik2020,2022_Rana,2023_Mishra}.

The European Space Agency's \textit{Euclid} mission promises to revolutionize weak-lensing science \citep{2025_Mellier}, as it surveys 14,000 square degrees of the extragalactic sky \citep{2022_Scaramella}. Its point-spread function (PSF) with full width at half maximum (FWHM) of about $0\farcs 16$ in the VIS band enables reliable measurements of the shapes of background galaxies, even when they are spatially close to the foreground lenses \citep{2021_Hoekstra}. As a result, galaxy-galaxy lensing analyses can be extended to projected scales of a few kiloparsecs, which is not possible with ground-based observations. By merging this exceptional high-resolution imaging dataset with spectroscopic redshifts from overlapping ground surveys, \textit{Euclid} will enable
high-fidelity galaxy-galaxy lensing measurements deep into the stellar mass regime.

Recent small-scale galaxy-galaxy lensing studies have investigated the matter distribution surrounding galaxies, using data from both the space-based Hubble Space Telescope and the ground-based Subaru Hyper Suprime-Cam survey \citep{2015_Kobayashi,2024_Khadka,2026_Fujikawa}. While these investigations highlighted the potential of galaxy-galaxy lensing for analysing the matter distribution on small scales, the smaller projected separations introduce observational systematics that are less consequential at larger scales. In particular, contamination from the light of the foreground lens can affect the measurement of background galaxy shapes, thereby potentially biasing the inferred lensing signal and matter profiles. Consequently, a comprehensive evaluation of these effects is essential to optimise the small-scale regime using next-generation imaging surveys, such as {\it Euclid}.

To this end, we explore the capabilities of {\it Euclid}-like small-scale galaxy-galaxy lensing measurements in conjunction with spectroscopic foreground samples to constrain total matter profiles. The structure of this paper is as follows.  \cref{sec:ggl} outlines the theoretical framework that governs our weak-lensing predictions. \cref{sec:data} discusses the methodology employed for constructing mock data. \cref{sec:img_sim} presents the simulation pipeline for images akin to those produced by the {\it Euclid} mission. Our primary findings are discussed in \cref{sec:results_n_discussions}, while a summary of the work is presented in \cref{sec:conclusions}.

Throughout this paper, we adopt a flat $\Lambda$CDM cosmology with the parameters $\Omega_{\mathrm{m}}=0.319$, $\Omega_{\mathrm{b}}=0.049$, $\sigma_8=0.813$, $n_{\mathrm{s}}=0.96$, and $h=0.67$, which corresponds to the reference cosmology utilized by \textit{Euclid} \citep{2025_castander}. All logarithmic calculations are conducted using base-10, and length scales are expressed in comoving units unless otherwise specified. Halo masses, denoted as $M_{\mathrm{200m}}$, are defined as the mass contained within a spherical radius that exhibits a mean interior density of 200 times the average matter density of the Universe.

\section{Galaxy galaxy lensing}\label{sec:ggl}

Weak gravitational lensing arises from the coherent distortion of background galaxy images caused by the differential bending of light by foreground mass distributions \citep[for recent reviews, see][]{2015_Kilbinger, 2018_Mandelbaum}. These distortions induce a shear in the observed galaxy ellipticities, preferentially in the tangential direction. The weak-lensing signal, $\Delta \Sigma$, for a given lens--source pair at redshifts $z_{\rm l}$ and $z_{\rm s}$, and at a lens-centric comoving separation $R$ with tangential shear $\gamma_{\rm t}$, is given by
\begin{equation}
    \Delta \Sigma(R)  = \Sigma_{\rm c}(z_{\rm l}, z_{\rm s}) \, \gamma_{\rm t},
\end{equation}
where $\Delta \Sigma(R) = \bar{\Sigma}(<R) - \langle \Sigma(R) \rangle$ is the excess surface density. Here, $\bar{\Sigma}(<R) = \int_0^R \Sigma(R') \, 2\pi R' \, \mathrm{d}R' / (\pi R^2)$ denotes the mean projected surface density within radius $R$, and $\langle \Sigma(R) \rangle$ is the azimuthally averaged projected surface density at radius $R$. The quantity $\Sigma_{\rm c}$ is the critical surface density, which characterises the geometric lensing efficiency of a lens--source pair and is given by
\begin{equation}
    \Sigma_{\rm c}(z_{\rm l}, z_{\rm s}) = \frac{c^2}{4\pi G} \frac{D_{\rm s}}{(1 + z_{\rm l})^2 D_{\rm l} D_{\rm ls}},
\end{equation}
where $D_{\rm l}$, $D_{\rm s}$, and $D_{\rm ls}$ are the angular diameter distances to the lens, to the source, and from the lens to the source, respectively. The factor $(1 + z_{\rm l})^2$ converts the physical critical surface density into comoving units. For lens--source pairs in which the source lies at or in front of the lens ($z_{\rm s} \leq z_{\rm l}$), the inverse critical surface density $\Sigma_{\rm c}^{-1}$ is set to zero, as these configurations do not contribute to the lensing signal.

To compute the lensing signal, we include the contributions from both the stellar and dark matter components. Although the gas component may include contributions from multiple phases, such as the cold interstellar medium and the hot X-ray halo gas, its overall mass is generally subdominant compared to the stellar and dark matter components for galaxies within our mass range \citep{2012_Papastergis}. For simplicity, we therefore neglect the gas component in this work.

For a lens with stellar mass $M_*$ and halo mass $M_{\rm h}$, the stellar contribution is modelled using a de~Vaucouleurs profile \citep{1948A_Dev}, with the half-light radius $R_{\rm e}(M_*)$ determined from the empirical relation of \citet{2019_Mowla}. The stellar surface density profile is
\begin{equation}
    \Sigma_{*}(R) = \Sigma_0 \exp\left[-b \left(\frac{R}{R_{\rm e}}\right)^{1/4}\right],
\end{equation}
where $b = 7.669$, and $\Sigma_0 = \frac{M_* b^8}{8 \pi R_{\rm e}^2 \Gamma(8)}$ is a normalisation constant fixed by the stellar mass of the lens. The dark matter contribution is modelled using an NFW  \citep{1997_Navarro}  profile with concentration parameter $c$ and halo mass $M_{\rm h}$, as described by the following equation:
\begin{eqnarray}
\rho(r) &=& \frac{\delta_{\rm c}\,\rho_{\rm m}}
{\left(r/r_{\rm s}\right)\left(1+r/r_{\rm s}\right)^2},\\
\delta_{\rm c} &=& \frac{200}{3}\,
\frac{c^3}{\ln(1+c)-c/(1+c)}, \qquad
r_{\rm s} = \frac{r_{\rm 200m}}{c},
\end{eqnarray}
where $\rho_{\rm m}$ is the average matter density of the universe and the corresponding projected surface density is computed using the expressions given in Eqs.~(11) and (13) of \citet{1999_Oaxaca}. We use these matter profiles to infer the equivalent radius, $R_{\rm eq}$, at which the stellar and dark matter contribute equally to the density.

\begin{figure}
    \centering
    \includegraphics[width=1.0\linewidth]{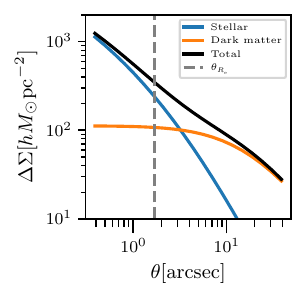}
\caption{Illustrative weak lensing signal as a function of angular separation $\theta$ for a for a galaxy with parameters at the median of our lens sample ($\log (M_* / h^{-2} M_\odot) = 10.43$,  $\log (M_{\rm h} / h^{-1} M_\odot) = 12.37$, $c=9.26$ and redshift $z=0.185$). The solid blue line indicates the stellar contribution, the solid orange line shows the dark matter component, and the solid black line denotes the total matter profile. The vertical grey dashed line marks the half-light radius, $\theta_{R_{\rm e}}$, for this galaxy, which is comparable to the seeing in ground-based observations, but considerably larger than the FWHM of the PSF of \textit{Euclid}.}
    \label{fig:dummy_plot}
\end{figure}

For reference, we show a typical weak-lensing signal as a function of angular separation, $\theta$, in \cref{fig:dummy_plot}. The profile is evaluated using the median properties of our lens sample (see \cref{sec:data} for details): a galaxy at redshift $z=0.185$ with a stellar mass of $\log (M_* / h^{-2} M_\odot) = 10.43$, a halo mass of $\log (M_h / h^{-1} M_\odot) = 12.37$, and a concentration of $c=9.26$. The solid blue, orange, and black lines show the stellar, dark matter, and total matter profiles, respectively. The vertical grey dashed line indicates the galaxy half-light radius, $\theta_{R_{\rm e}}$.

\Cref{fig:dummy_plot} highlights the challenge of measuring the weak lensing at scales where the stellar mass dominates the lensing signal. The half-light radius of the lens is comparable to the seeing in ground-based surveys. However, as shown in \cite{2021_Hoekstra}, even measurements using \textit{Euclid} data will be affected by blending for $\theta<3''$. It is therefore essential to quantify the impact of (residual) lens light on our ability to measure the lensing signal down to these small scales. 

For our lens sample, source galaxy positions are sampled within a comoving projected separation of $100 \, h^{-1} \, \mathrm{kpc}$, assuming the lenses themselves are isolated systems. The expected shear $\gamma$ is applied to the intrinsic complex ellipticities $\epsilon_{\rm s}$ following the prescription of \citet{1997_Seitz}, producing the observed ellipticities $\epsilon$:
\begin{equation}
\epsilon = 
\begin{cases}
    \dfrac{\epsilon_{\rm s} + g}{1 + g^* \epsilon_{\rm s}}, & |g| < 1, \\[10pt]
    \dfrac{1 + g \epsilon_{\rm s}^*}{\epsilon_{\rm s}^* + g}, & |g| > 1,
\end{cases}
\end{equation}
where the reduced shear $g = \gamma / (1 - \kappa)$ is complex, and $\kappa = \Sigma / \Sigma_{\rm c}$ is the convergence. An asterisk denotes complex conjugation. 

Because the shear signal from an individual source galaxy is typically small compared to its intrinsic ellipticity, we measure the weak-lensing signal by stacking background sources around the lens galaxies, which increases the signal-to-noise ratio because the intrinsic ellipticities of the source galaxies are randomly oriented and therefore average to zero. 

We measure the stacked weak-lensing signal $\Delta \Sigma (R_{\rm p})$ for each of our data bins using fourteen logarithmically spaced comoving projected radial bins spanning $6$–$100 \, h^{-1} \, \mathrm{kpc}$. We adopt the maximum-likelihood estimator of \citet{2004_Sheldon}, expressed as
\begin{equation}
    \Delta \Sigma (R_{\rm p}) = \frac{\sum_{{\rm ls} \in R_{\rm p}} w_{\rm ls} \, \epsilon_{\rm t} \, \Sigma_{\rm c}(z_{\rm l}, z_{\rm s})}{\sum_{{\rm ls} \in R_{\rm p}} w_{\rm ls}},
\end{equation}
where the weight for each lens--source pair is $w_{\rm ls} = w_{\rm l} w_{\rm s} \Sigma_{\rm c}^{-2}$. Here, $w_{\rm l}$ and $w_{\rm s}$ are the weights assigned to the lens and source galaxies, respectively, and $\epsilon_{\rm t}$ is the tangential ellipticity component relative to the lens. In our mock catalogues, we assume unit weights, $w_{\rm l} = w_{\rm s} = 1$, along with a conservative source selection to reduce potential contamination from physically associated sources and thereby mitigate the impact of intrinsic alignments. To ensure a clean background source sample, we require $z_{\rm s} > z_{\rm l, max} + z_{\rm diff}$, where $z_{\rm l, max} = 0.4$ is the maximum lens redshift, adding a small buffer of $z_{\rm diff} = 0.2$. To mitigate shape noise in our mock measurements and to recover a smooth underlying signal, we artificially boost the signal-to-noise ratio by using a lens sample that is $100$ times larger than the expected survey yield. However, to ensure that our parameter inference accurately reflects the realistic constraining power of the survey, the statistical uncertainties are calculated using the true expected number of lenses. We compute the covariance matrix of our measurements using the jackknife resampling technique \citep{1974_miller} with one hundred jackknife samples. As our measurements are within the shape-noise dominated regime, the off-diagonal elements are negligible; consequently, we only consider the diagonal terms of the covariance matrix in this study.

Since our focus is on small scales, we consider the reduced shear correction in the regime where $\kappa < 1$ \citep[e.g.,][]{McClintock2018, Pereira2020, Umetsu2020} as
\begin{equation}
    \Delta \Sigma \rightarrow \Delta \Sigma \left( 1 + \sum_{n=1}^5 \Sigma^n \, \langle \Sigma_{\rm c}^{-n} \rangle\right) \, ,
\end{equation}
where $\langle \Sigma_{\rm c}^{-n} \rangle$ represents the average inverse critical surface density across the lens and source redshift distributions. We expand our model up to the fifth order to ensure that any variations in our predictions from higher-order terms remain well within the statistical uncertainties of our signal measurements. 

Our model assumes that the stellar component follows stacked de Vaucouleurs, scaled by the mismatch parameter $\alpha =M_*^{\rm true}/M_*^{\rm obs}$, which is the ratio between true stellar mass $M_*^{\rm true}$ and the observed stellar mass $M_*^{\rm obs}$. The dark matter contribution is described by the NFW profile. Using a generalized NFW profile \citet{2001_Wyithe} instead, we verified that our adopted length scales are insensitive to the inner dark matter slope, similar to past studies \citep[e.g.,][]{2006_Mandelbaum, 2013_Newman}.

Our model therefore contains three parameters: the mismatch parameter $\alpha$, the halo mass $\log(M_{\rm h}/ h^{-1} M_\odot)$, and the concentration $c$. The model components are expressed as
\begin{align}
    \Delta\Sigma &= \Delta\Sigma_{\rm dm} + \alpha \langle \Delta\Sigma_{\rm *} \rangle, \\
    \Sigma &= \Sigma_{\rm dm} + \alpha \langle \Sigma_{\rm *} \rangle,
\end{align}
where the subscripts ${\rm dm}$ and ${\rm *}$ refer to the dark matter and stellar components, respectively. The angle brackets denote quantities that are averaged over each stellar mass bin. 

We adopt a Gaussian likelihood $\mathcal{P}(\mathcal{D}|\Theta)$ for the Bayesian inference of the model parameters $\Theta$, and the corresponding posterior $\mathcal{P}(\Theta | \mathcal{D})$ is given by Bayes theorem:
\begin{equation}
    \mathcal{P}(\Theta | \mathcal{D}) \propto \mathcal{P}(\mathcal{D}|\Theta) \, \mathcal{P}(\Theta),
\end{equation}
with
\begin{equation}
    \mathcal{P}(\mathcal{D}|\Theta) \propto \exp(- \chi^2 / 2),
\end{equation}
where $\chi^2 = [\mathcal{D} - M(\Theta)]^{\mathrm{T}} C^{-1} [\mathcal{D} - M(\Theta)] $, $\mathcal{D} $ represents the data vector, and $ M(\Theta) $ denotes the model prediction vector for the parameter set $ \Theta = [\alpha, \log( M_{\mathrm{h}}/ h^{-1} M_\odot), c] $. We adopt uniform priors for $ \alpha $ in the range of 0.1 to 5, for $ \log (M_{\mathrm{h}} / h^{-1} M_\odot) $ from 9 to 16, and for $ c $ from 0 to 20.

To mitigate the effect of noise in the covariance matrix, which could lead to biased inferences, we follow Eq. (17) of \citet{2007_Hartlap} and apply a correction to the inverse of the covariance matrix. We utilize the affine-invariant Markov Chain Monte Carlo (MCMC) sampler \citep{2010_Goodman}, as implemented in the \texttt{emcee} Python package \citep{2013_Foreman} for inferring the posterior distributions of the model parameters. For each fit, we use 64 walkers, discarding the first 2,000 steps as burn-in, and we perform an additional 4,000 steps per walker to sample the posterior distribution.

Our main analysis involves running independent MCMC chains for each of the 27 stellar mass bins. Lastly, we repeat this exact inference procedure across three distinct redshift intervals at fixed stellar mass to explicitly investigate the redshift evolution of the  mismatch parameter,  $\alpha $.

\section{Input catalogues}
\label{sec:data}

In this section, we describe the simulated lens and source galaxy datasets used in our analysis, along with the applied selection cuts and the expected survey configuration.

\subsection{Lenses}
\label{sec:lenses}

We construct the lens sample using the public \textsc{Flagship} mock galaxy catalogue  \citep{2025_castander}. The simulated volume is comparable to that expected for the \textit{Euclid} survey and resolves approximately 16 billion dark matter haloes, identified using the Rockstar halo finder \citep{2013_behroozi}. The galaxy catalogue is generated by combining halo occupation distribution (HOD; e.g., \citealp{1998_Jing,2000_Peacock,2002_Berlind}) and subhalo abundance matching (SHAM; e.g., \citealp{2004_Vale,2006_Conroy,2010_Guo}) techniques with empirical prescriptions for galaxy properties \citep{2025_castander}. For each galaxy, the catalogue provides central galaxy selection flags, three-dimensional positions, redshifts, broadband photometry, stellar masses, and host halo properties.

\begin{figure}
\centering
\includegraphics[width=\linewidth]{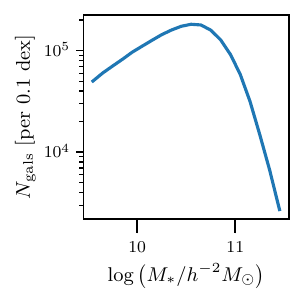}
\caption{The number of central lens galaxies as a function of stellar mass for the mock DESI BGS sample used in this work. The histogram is scaled to the expected DESI--\textit{Euclid} overlap area of approximately $9000 \, \mathrm{deg}^2$.}
\label{fig:lens_smdist}
\end{figure}

We use the central flag to select central galaxies from the mock catalogue that have an apparent magnitude of \(m_r < 19.5\) and fall within the redshift range \(0.0 < z < 0.4\). This selection is representative of the Dark Energy Spectroscopic Instrument Bright Galaxy Survey (DESI BGS; \citealp{2023_Hahn}), which is anticipated to overlap with the \textit{Euclid} survey over an area of approximately \(9000\,\mathrm{deg}^2\) \citep{2023_Naidoo}. Since the public \textsc{Flagship} release only covers one-eighth of the full sky, we scale the lens sample to align with the expected overlap area between DESI and \textit{Euclid}. For each selected galaxy, we extract several parameters, including stellar mass, halo mass, halo concentration, DES \(r\)-band magnitude, and \textit{Euclid} VIS-band magnitude, as provided in the catalogue. We convert the provided virial halo masses and the associated concentrations to the \(M_{\rm 200m}\) spherical over-density definition using the public \texttt{COLOSSUS} package \citep{2015_Diemer_colossus}.

Our fiducial analysis assumes a sample of only central galaxies that are considered to be effectively infinitely spatially separated from one another. These galaxies are further divided into 27 stellar mass bins, and \cref{fig:lens_smdist} shows the distribution of galaxies across these logarithmic stellar mass bins. To study the redshift evolution of $\alpha$, we also defined a subsample of lenses limited to a stellar mass range of \(10.5 < \log(M_* / h^{-2} M_\odot) < 11.0\). This subsample is split into three redshift intervals: \(0.0 < z < 0.1\), \(0.1 < z < 0.2\), and \(0.2 < z < 0.3\). We also check the impact of the isolation assumption and the inclusion of satellite galaxies on our results. In \cref{subsec:satellites_galaxies}, we examine the impact of satellite galaxies using an alternate approach. Also, we follow the method outlined by \citet{2017_Wright} to apply a color-homogeneity selection, which helps to eliminate the color excess biases present in our lens sample. 

\begin{figure}
    \centering    \includegraphics[width=1.0\linewidth]{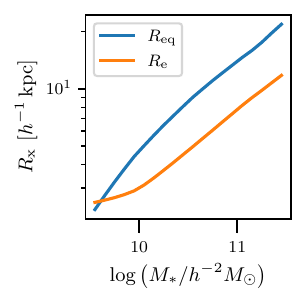}
    \caption{Comparison between the median half-light radius and the equivalent radius as a function of stellar mass.
    The solid blue curve indicates the projected separation at which the baryonic and dark matter contributions are equal ($R_{\rm eq}$), while the solid orange curve shows the half-light radius ($R_{\rm e}$) of the galaxies.}
    \label{fig:re_req}
\end{figure}

While the \textsc{Flagship} catalogue also provides sheared background source galaxies around each of our lenses for weak-lensing studies, we do not utilize the simulated shear directly. The angular scales examined in our study focus on the innermost regions of galaxy haloes, where the spatial resolution of the \textsc{Flagship} lensing maps is inadequate for our analysis. Instead, we generate the weak-lensing signal using the analytical framework described in \cref{sec:ggl}.
Therefore, the projected matter distribution surrounding each lens is given by the sum of a de Vaucouleurs profile, which describes the stellar component, and an NFW profile that represents the dark matter halo. In \cref{fig:re_req}, we compare the half-light radius ($R_{\rm e}$, solid orange line) and the equivalent radius ($R_{\rm eq}$, solid blue line). Our findings indicate that $R_{\rm eq}$ is larger than $R_{\rm e}$ across most stellar mass bins, suggesting that the projected length scales dominated by stellar mass are suitable for small-scale lensing studies.

\subsubsection{Satellite galaxies and isolation criteria}\label{subsec:satellites_galaxies}
In our fiducial analysis, we focus exclusively on central galaxies within the lens sample to generate the mock data vectors. However, in observational studies, satellite galaxies also contribute to the measured weak lensing signals, particularly on the small scales observed in our galaxy-galaxy lensing measurements. In addition to being part of larger host halos, satellite galaxies experience environmental processes such as tidal interactions and ram-pressure stripping. These processes can alter their baryonic properties and introduce additional dependencies that go beyond the stellar mass-halo mass relation \citep[e.g.,][]{2026_Yin}.

Identifying central galaxies through observation is a challenging task because it requires an understanding of the underlying halo occupation and assembly history. The presence of satellite galaxies adds complexity due to the matter content of their own subhalos and the larger host halo, which can lead to biased interpretations if not properly addressed. 

To minimize this contamination, we  can apply a strict isolation criterion to the mock DESI BGS sample. Specifically, we remove any galaxy that has a more massive neighbor within a projected comoving separation of $300 \,h^{-1}\mathrm{kpc}$ at the lens redshift. This isolation scale is three times larger than the maximum projected separation considered in our lensing analysis. As a result, this selection process preferentially removes galaxies that are part of larger host systems, thereby producing a sample that is predominantly composed of central galaxies.

\begin{figure}
\centering
\includegraphics[width=\linewidth]{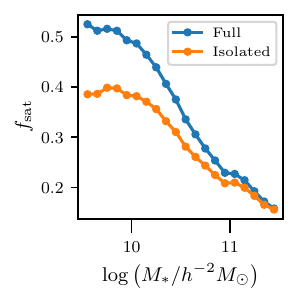}
\caption{Effect of isolation criteria on the satellite fraction. The blue points show the satellite fraction as a function of stellar mass for the full lens sample. The orange points show the remaining satellite fraction after applying the isolation criterion described in the text.}
\label{fig:sat_fractions}
\end{figure}

\cref{fig:sat_fractions} shows the impact of this isolation criterion on the satellite fraction. The blue points represent the satellite fraction before applying the selection, while the orange points show the remaining fraction after isolation. The reduction is strongest in the lower stellar mass bins. This trend is expected because lower-mass galaxies are more likely to reside as satellites within the extended environments of more massive systems. Furthermore, the lower stellar mass bins are preferentially located at lower redshifts, where a fixed physical isolation scale corresponds to a larger angular separation, increasing the effectiveness of the projected neighbour selection. 

Although the isolation criterion substantially reduces the contribution from satellite galaxies, a significant satellite population remains. In addition to their central galaxies, the subhalos of these satellites contribute to the expected weak-lensing signal. Since the \textsc{Flagship} simulation does not explicitly provide subhalo masses, we model the contribution from these remaining satellites following the methodology of \citet{2015_Zu}. We assign halo masses and concentrations to satellite galaxies by matching them to central galaxies with similar stellar masses ($\Delta \log M_* \leq 0.01$ dex) and redshifts ($\Delta z \leq 0.01$). We model the dark-matter distribution of these subhalos using a truncated NFW profile to account for tidal stripping and following \citet{2015_Zu} we conservatively truncate the satellite halo profiles at $0.4R_{\rm 200m}$.

This procedure provides an efficient approach for incorporating satellite contributions into our mock lensing analysis. But in typical observational studies, these contributions are typically modelled statistically through a halo model that simultaneously describes the lensing signal from central and satellite galaxies \citep[e.g.,][]{2013_Vandenbosch, 2023_Mishra, 2026_Fujikawa}. While such an approach provides a more complete description of the galaxy--matter connection, it also introduces additional degrees of freedom into the inference framework. Most importantly, the degeneracies between the satellite halo occupation parameters and the stellar contributions may reduce the constraints on the mismatch parameter $\alpha$. To fully assess the impact of such an approach, we require higher-resolution ray tracing simulations.
Nonetheless, we expect that our results yield a fair estimate of what a more sophisticated analysis would yield.

As noted above, satellite galaxies preferentially reside in dense environments, which introduces additional complexity from correlated structures \citep[][]{2016_Fang}. Beyond the physical modelling of satellite contributions, realistic observations also require accounting for image-level effects arising from complex galaxy environments. Source galaxies located near clustered satellite systems and their associated central galaxies can be affected by overlapping galaxy light, blending, and crowding, which may bias source detection and shape measurements. These complex effects require dedicated image simulations tailored to survey conditions, including realistic galaxy populations, light profiles, clustered satellites, and weak lensing distortions \citep[e.g.,][]{2015_Sifon}.

Our alternative approach can bypass both the modelling and observational complexities. We apply the isolation criterion described above and assess its impact on the constraints on $\alpha$. As discussed in \cref{appex:alpha}, the resulting constraints differ only marginally from those obtained using the only central-galaxy fiducial lens sample, indicating our treatment for the satellite contribution does not significantly affect our fiducial results. The isolation criterion therefore provides a simple means of reducing potential satellite contamination without requiring explicit modelling of the satellite population.

\subsection{Sources}
\label{sec:sources}

To simulate the galaxy--galaxy lensing measurements, we construct a mock source catalogue designed to reproduce the expected characteristics of a \textit{Euclid}-like weak-lensing survey. We assume a mean source density of $30 \, \mathrm{arcmin}^{-2}$ and adopt the analytical source redshift distribution, $n(z_{\rm src})$, given by Eq.~(113) of \citet{2020_Blanchard}. The resulting normalised redshift distribution has a median source redshift of approximately $z_{\rm med} \simeq 0.8$, consistent with the expected depth of the \textit{Euclid} survey.

For each lens galaxy, we populate the surrounding field by drawing the number of background sources from a Poisson distribution with a mean source density and the simulated image area. The source positions are then sampled uniformly across the field of view with the associated source redshift is assigned by randomly sampling from the $n(z_{\rm src})$ and an intrinsic ellipticity, $\epsilon_{\rm s}$, by drawing independently for each ellipticity component from a Gaussian distribution with zero mean and a root-mean-square dispersion of $\sigma_\epsilon = 0.26$ \citep{2011_Laureijs}. This intrinsic shape distribution reproduces the level of shape noise expected for a \textit{Euclid}-like survey.

The observed ellipticity of each source is obtained by evaluating the reduced shear, $g$, at its projected position relative to every foreground lens and combining it with the intrinsic ellipticity using the standard weak-lensing formalism described in \cref{sec:ggl}. To ensure that our analysis remains within the weak-lensing regime, we exclude source galaxies for which the convergence or reduced shear exceeds the limits $\kappa < 1.0$ and $|g| < 0.3$.

\section{Image simulations}\label{sec:img_sim}

On scales of approximately \(10\,h^{-1}\) kpc, small-scale weak lensing measurements can be affected by light from foreground lenses. This contamination distorts the observed shapes of background source galaxies, leading to biases in the measured galaxy-galaxy lensing signal and in the inferred model parameters. In this section, we utilize large-scale image simulations based on a survey configuration similar to that of \textit{Euclid} to assess the impact of this systematic effect.

We generate our image simulations using the publicly available Python package \textsc{GalSim} \citep{2015_Rowe}, which has been extensively validated for weak-lensing studies during the GREAT3 challenge \citep{2014_Mandelbaumm, 2015_Mandelbaumm}. Following  \citet{2017_Hoekstra} and \citet{2021_Hoekstra}, we model the PSF of the \textit{Euclid} telescope in its \textit{VIS} band using an Airy profile. This model assumes a primary mirror diameter of $1.2\,\mathrm{m}$, a central obscuration of $0.3$, and a reference wavelength of $800\,\mathrm{nm}$ \citep{Cropper18}. 
We set the Gaussian noise level such that galaxies with a full FWHM of $0\farcs3$ and a magnitude of $I_{\rm E}=25$ are detected with a signal-to-noise ratio of approximately $10$ \citep{2025_Mellier}, as measured by \textsc{SExtractor} \citep{1996_Bertin}.

For each lens, we assume a de Vaucouleurs light profile, along with physical parameters described in \cref{sec:ggl} and \cref{sec:lenses}. Using these profiles, we generate postage-stamp images matching the \textit{Euclid} pixel scale of $0\farcs1$, with the stamp size determined by the maximum projected radial bin and the lens redshift. For the background source population, we adopt a simplified but conservative setup by assuming all galaxies lie at the \textit{Euclid} detection limit. Specifically, each source galaxy is modelled using a Gaussian profile with an FWHM of $0\farcs3$ and a fixed magnitude of $I_{\rm E}=25$, while other properties are drawn as described in \cref{sec:sources}.
These faint sources dominate the overall source number density and are a major contributor to the weak lensing measurements. We distribute sources spatially randomly around the lenses, with the exact number of sources per stamp drawn from a Poisson distribution with a mean tied to this average density.

\begin{table}
    \centering
    \begin{tabular}{ll}
    \hline
    \texttt{DETECT\_MINAREA} & 6\\
    \texttt{DETECT\_THRESH} & 1.5\\
    \texttt{FILTER\_NAME} & \texttt{gauss\_3.0\_7$\times$7.conv}\\
    \texttt{DEBLEND\_NTHRESH} & 32\\
    \texttt{DEBLEND\_MINCONT} & 0.005\\
    \texttt{CLEAN\_PARAM} & 1.0\\
    \texttt{BACK\_SIZE} & 128\\
    \texttt{BACK\_FILTERSIZE} & 3\\
    \hline
    \end{tabular}
    \caption{\textit{\textsc{SExtractor} configuration parameters.} The table above describes the key \textsc{SExtractor} parameters adopted for source detection and deblending in the simulated images. The configuration defines the detection threshold, minimum detection area, convolution filter, deblending settings, and background estimation used to generate the source catalogue for the subsequent weak-lensing analysis.}
    \label{Table:1}
\end{table}

Source detection is performed using \textsc{SExtractor} \citep{1996_Bertin} with the typical parameter values listed in \cref{Table:1}, closely following the methodology of \citet{2021_Hoekstra}. In our implementation, \textsc{SExtractor} internally estimates the background level, which is subsequently used in conjunction with the \texttt{DETECT\_MINAREA} and \texttt{DETECT\_THRESH} parameters to identify objects. The detected sources are stored in a catalogue and passed to the shape measurement pipeline. To quantify biases in the weak-lensing signal, we measure the shapes of these detected source galaxies using the KSB algorithm \citep{1995_Kaiser, 1998_Hoekstra}, adopting the specific implementation described by \citet{2017_Hoekstra}. Because the \textit{Euclid} pixel scale is relatively large compared to the diffraction limit of the telescope, insufficient PSF sampling could introduce artificial biases into the shape measurements. To mitigate this, we follow \citet{2021_Hoekstra} and oversample the PSF by a factor of five, which significantly improves the PSF characterisation. Furthermore, we apply a Gaussian weight function with a width of three pixels during the shape measurement process.

\begin{figure}
    \centering
    \includegraphics[width=\columnwidth]{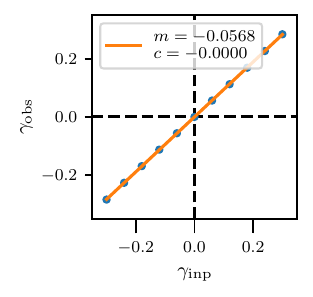}
    \caption{The blue data points show the recovered shear measurements, while the solid orange line represents the best-fit linear model described by \cref{eq:shecal}. The slope corresponds to the multiplicative shear bias $m$.}
    \label{fig:shecal}
\end{figure}

To accurately derive the shear calibration biases, we separately generated a dedicated set of image simulations containing exclusively background source galaxies, constructed on extensive grids of $1550 \times 1550$ pixels. We produced a comprehensive ensemble of these simulated images across eleven discrete, known input shears, $\gamma_{\rm inp}$, which were uniformly distributed between $-0.3$ and $0.3$. The observed shear, $\gamma_{\rm obs}$, is subsequently computed from the ensemble-averaged shape estimate, $\langle \chi \rangle$, and the pre-seeing shear polarisability, $\langle P^\gamma \rangle$ \citep{Luppino1997}, according to the relation:
\begin{equation}
    \gamma_{\rm obs} = \frac{\langle \chi \rangle}{\langle P^\gamma \rangle},
\end{equation}
where the ensemble average is evaluated using uniform unit source weights. The choice of unit weights is motivated by the standard observational analyses, where inverse-variance weights are typically a function of specific source population properties, such as galaxy magnitudes \citep[see, e.g.,][]{2023_Li_shun-sheng, 2022PASJ_Li_Xiangchong, 2018_Mandelbaum_hsc_y1}. However, because our idealised simulations consider only a single, uniform population of source galaxies, we apply equal statistical weight to each source.

We model the shear response using a standard linear calibration formalism:
\begin{equation}
    \gamma_{\rm obs} = (1+m)\,\gamma_{\rm inp} + c,
    \label{eq:shecal}
\end{equation}
where $m$ and $c$ parameterise the multiplicative and additive biases, respectively. The optimal best-fit parameters are then determined via a Levenberg-Marquardt least-squares minimisation algorithm \citep{1981_Gill}. 

\cref{fig:shecal} shows the resulting shear calibration relation. The blue data points represent the discrete shear measurements recovered directly from the image simulations, while the solid orange line denotes the best-fit linear model. From this analysis, we derive a multiplicative bias of $m = -0.0568$ and an additive bias of $c = 0.0$. We systematically apply these derived calibration factors to all subsequent measurements in our analysis, ensuring unbiased shear estimates around the simulated lenses.

\begin{table}
    \centering
    \begin{tabular}{ll}
    \hline
    Centroid positions & $x,y$\\
    Integrated magnitude & $m$\\
    Sérsic index  & $n$  \\
    Half-light radius & $r_{\rm e}$ \\ 
    Position angle & $\phi = 0$ (Fixed)\\ 
    Axis ratio & $q = 1$ (Fixed)\\ 
    \hline
    \end{tabular}
    \caption{\textit{\textsc{Galfit} model parameters.} The table above represents the free and fixed parameters used to model the lens galaxy light distribution with a single Sérsic profile in \textsc{Galfit}. The centroid position ($x,y$), integrated magnitude ($m$), Sérsic index ($n$), and effective (half-light) radius ($r_{\rm e}$) are treated as free parameters, while the position angle ($\phi$) and axis ratio ($q$) are fixed to 0 and 1, respectively. The best-fitting model is subtracted from the science image to produce the lens-light-subtracted images used in the subsequent weak-lensing analysis.}
    \label{galfit_params}
\end{table}

To actively mitigate contamination from foreground lens light, we explore an image-subtraction approach. Similar to our pure-source calibration suite, we run a large batch of image simulations containing both foreground lenses and background sources to ensure a statistically robust evaluation of our mitigation strategy. We model the lens light profiles using \textsc{Galfit} \citep{2002_Peng}, a robust two-dimensional fitting algorithm capable of accommodating a variety of surface brightness profiles. As detailed in \cref{galfit_params}, we adopt a Sérsic profile with a fixed position angle $\phi = 0$ and axis ratio $q = 1$. The profile is further characterised by its Sérsic index $n$, half-light radius $r_{\mathrm{e}}$, integrated magnitude $m$, and centroid coordinates $(x,y)$. Initial parameter guesses for the position, magnitude, and half-light radius are supplied directly from the \textsc{SExtractor} catalogue, while the initial Sérsic index guess value is set to $n = 4$. We fix the model fitting region to ten times the half-light radius reported by \textsc{SExtractor}. 

Our \textsc{Galfit} procedure processes the simulated images using the oversampled point spread function (PSF) and a sigma image derived from the intrinsic noise characteristics of the simulations. While more complex profiles can be applied to individual lenses if necessary, we opt for \textsc{Galfit} because of its widespread use and computational efficiency in localized postage-stamp simulations. However, we recognize that in full survey pipelines, highly optimized multi-band deblenders like \textsc{Scarlet} \citep{2018_Melchior} or forward-modeling frameworks such as \textsc{The Tractor} \citep{2016_Lang} are typically employed to differentiate between foreground and background emissions. Additionally, more sophisticated forward modeling approaches, like \textsc{LensMC} \citep{2024_Congedo}, can integrate similar modeling within a complete image simulation framework.

Once the best-fit lens light models are subtracted, the residual images are reprocessed through the \textsc{SExtractor} and KSB pipelines to generate updated shape catalogues. We then use these updated catalogues to assess the efficacy of lens light subtraction on the recovery of the true weak-lensing signal.

\subsection{Lens light systematics}\label{sec:lens_light_img_sim}

\begin{figure}
    \centering
    \includegraphics[width=\linewidth]{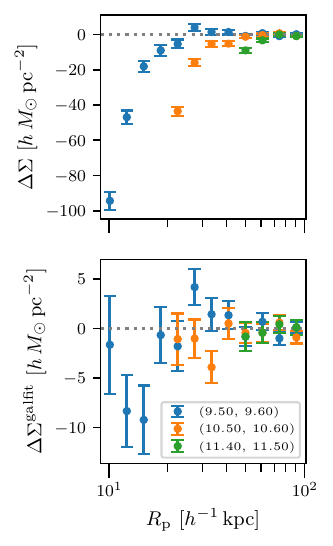}
    \caption{\textit{Top panel:} Bias in the measured $\Delta\Sigma$ signal when ignoring the contamination by foreground lens light. \textit{Bottom panel:} the residual systematic after modelling and subtracting the foreground lens light using \textsc{Galfit}. The blue, orange, and green points correspond to lens galaxies selected in the stellar-mass ranges $9.5 < \log(M_*/h^{-2} M_\odot) < 9.6$, $10.5 < \log(M_*/h^{-2} M_\odot) < 10.6$, and $11.4 < \log(M_*/h^{-2} M_\odot) < 11.5$, respectively. The error bars represent jackknife uncertainties.}
    \label{fig:esd_ll_sys}
\end{figure}

We assess the impact of the extended surface-brightness distribution of the lens galaxy on both the detection and shape measurement of background sources across three representative stellar mass ranges: $9.5 < \log(M_*/h^{-2} M_\odot) < 9.6$, $10.5 < \log(M_*/h^{-2} M_\odot) < 10.6$, and $11.4 < \log(M_*/h^{-2} M_\odot) < 11.5$. These intervals correspond to the low, intermediate, and high mass regimes of our lens sample, enabling us to evaluate how lens-light contamination depends on galaxy luminosity and size. We simulate background sources without any applied gravitational shear, ensuring that any measured signal directly reflects the systematic contribution induced by the foreground lens light. The shapes are then measured using the KSB algorithm (see \cref{sec:img_sim}) without applying lens-light subtraction or blending corrections, as our goal is to quantify the raw systematics.

The top panel of \cref{fig:esd_ll_sys} shows the resulting bias in the measured $\Delta\Sigma$. The foreground lens light artificially stretches the measured source ellipticities along the lens-source axis. This induced radial alignment mimics a negative tangential shear, which directly suppresses the $\Delta\Sigma$ signal, consistent with previous studies \citep{1998_Hudson,2015_Sifon}. The amplitude of this bias decreases with increasing projected separation, reflecting the diminishing surface brightness of the foreground lens at larger radii. The systematic also exhibits a clear dependence on stellar mass, as more massive lens galaxies produce larger biases due to their brighter and more extended light profiles.
Interestingly, we find that the profiles are very similar when plotted in units of the half-light radius of the lens. This motivates our cut in terms of $R_{\rm e}$ later on. Similar radial and stellar-mass dependencies are observed in the stacked source photometry and source-detection fractions discussed below, indicating a shared contamination mechanism driven by the foreground light distribution. 

For the lens populations considered in this analysis, the induced systematic signal can reach an amplitude comparable to the expected galaxy--galaxy lensing signal itself (\cref{fig:esd_iso_cen}). This is most severe for the lowest-mass interval ($9.5 < \log(M_*/h^{-2} M_\odot) < 9.6$), where the bias competes with the measured signal at the smallest projected scales, highlighting the critical need to accurately model foreground lens light when extracting small-scale lensing constraints and interpreting the connection between the stellar and dark matter components of galaxies. We note that similar blending and shear biases are critical considerations in contemporary Stage-III surveys \citep[e.g.,][]{2018_Mandelbaum, 2022_MacCrann}.

We investigate the effectiveness of model-based lens-light subtraction in mitigating this systematic effect. Using \textsc{Galfit}, we model the foreground lens light and subtract the best-fitting galaxy model before repeating the source shape measurements. The resulting $\Delta\Sigma$ measurements are presented in the bottom panel of \cref{fig:esd_ll_sys}. Lens-light subtraction reduces the systematic bias by approximately an order of magnitude, leaving residuals well below the expected statistical uncertainties for these measurements. However, the effectiveness of this correction depends on the efficiency of the deblending algorithm in accurately recovering the intrinsic source light distributions and shapes. The magnitude of the uncorrected bias demonstrates the importance of validating subtraction methods using realistic image simulations; when lens-light contamination approaches the amplitude of the true lensing signal, residual imperfections in the foreground light model can propagate into the inferred shear signal, consequently biasing parameters such as the stellar mismatch parameter, $\alpha$.

Foreground lens light affects not only the measurement of source shapes but also the detection efficiency of background galaxies and biases their flux measurements. These introduce additional scale-dependent systematic uncertainties. These are explored in detail in \cref{app:blending}. Although their impact is less prominent than the bias in the shape estimation, careful removal of lens light can largely account for them. 

After applying lens-light modelling and subtraction, the scale-dependent detection bias is substantially reduced, and the recovered source population closely matches that obtained in the absence of foreground lens contamination. However, our simulations are idealised and do not include more realistic survey conditions, such as complex galaxy environments, overlapping light profiles, and source blending. These effects will become increasingly relevant in dense environments (e.g. galaxy groups and clusters), where multiple bright galaxies and intracluster light further complicate source detection, photometry, and shape estimation. 

Foreground lens light therefore remains an image-level systematic that requires careful validation before physical constraints are extracted from small-scale galaxy--galaxy lensing measurements. A comprehensive assessment using full \textit{Euclid}-like image simulations and forward-modelling techniques will be presented in future work.

\section{Results and Discussions} \label{sec:results_n_discussions}
In this section, we present our results from the mock galaxy-galaxy lensing measurement on the simulated DESI-BGS spectroscopic galaxies. As discussed in \cref{sec:ggl}, we consider only central galaxies and treat them as perfectly isolated lenses free from the projection effects of neighbouring structures, as our fiducial analysis. The background source galaxy population is modelled to precisely match the expected observational properties of the upcoming \textit{Euclid} survey. 

\subsection{Lensing Measurements}\label{sec:meas_lensing}
\begin{figure}
    \centering
    \includegraphics[width=\linewidth]{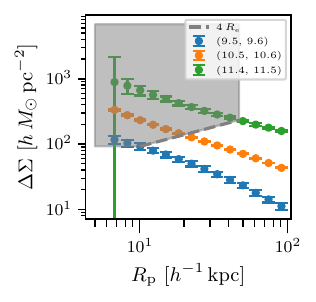}
    \caption{Galaxy--galaxy lensing signal, $\Delta\Sigma$, as a function of radius around central galaxies in three stellar mass bins, indicated by the blue, orange, and green data points, shown with their corresponding jackknife errors. The grey-shaded region, bounded by the dashed line, represents projected radial scales smaller than four times the median half-light radius of the sample; these inner scales are excluded from the analysis in this study.}
    \label{fig:esd_iso_cen}
\end{figure}

In \cref{fig:esd_iso_cen}, we show the measured weak lensing signal $\Delta\Sigma$ in fourteen logarithmically spaced projected comoving radial bins in the range from $10 \, h^{-1} \mathrm{kpc}$ to $100 \, h^{-1} \mathrm{kpc}$ for lenses divided into three representative stellar mass bins. The blue, orange, and green data points represent the intrinsic, noiseless lensing signals for galaxies within the stellar mass intervals of $9.5 < \log(M_*/h^{-2} M_\odot) < 9.6$, $10.5 < \log(M_*/h^{-2} M_\odot) < 10.6$, and $11.4 < \log(M_*/h^{-2} M_\odot) < 11.5$, respectively along with the error bars representing statistical uncertainties. 

The induced weak lensing shear is directly proportional to the projected matter distribution of the foreground lens. As expected, the signal exhibits a monotonic decrease as a function of the projected radial distance from the galaxy centre. A comparison of the lensing signal amplitudes across the distinct mass bins reveals a systematically increase in the amplitude with stellar mass. This trend is in  agreement with expectations from previous observational studies \citep[for e.g.][]{2015_Zu, Dvornik2020, 2023_Mishra}. 

At very small projected separations, the light from the foreground lens significantly influences the information content of the background sources. This effect becomes increasingly important as we study more massive galaxies. To mitigate these photometric and blending effects, we adopt an inner radial cut motivated by our image simulations  (\cref{sec:lens_light_img_sim}). This cut excludes all measurements at scales smaller than four times the median half-light radius of the lenses within each respective stellar mass bin. This exclusion is shown by the grey-shaded region bounded by a dashed line in \cref{fig:esd_iso_cen}. The corresponding signal-to-noise ratios (S/N) for these measurements can be as large as roughly 150, as shown in \cref{fig:snr_iso_cen}.

\subsection{Mismatch parameter constraints}
\begin{figure*}
\centering
\includegraphics[width=\textwidth]{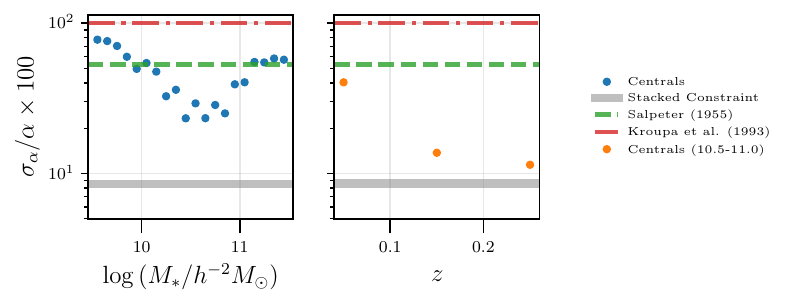}
\caption{\textit{Left panel:} Relative uncertainty in the stellar mismatch parameter, $\alpha$, as a function of stellar mass (blue points). The lens sample is divided into 0.1 dex stellar mass bins spanning $9.5 \leq \log(M_*/h^{-2} M_\odot) \leq 11.5$. The expected offsets relative to a fiducial Chabrier IMF for Salpeter and Kroupa IMFs are shown by the green dashed and red dot-dashed lines, respectively. \textit{Right panel:} Similar, but for centrals with stellar mass
$10.5 \leq \log(M_*/h^{-2} M_\odot) \leq 11$ as a function of redshift.}
\label{fig:alpha_constraints}
\end{figure*}

We model the weak lensing measurements presented in \cref{sec:meas_lensing} using the analytical framework described in \cref{sec:ggl}. The galaxy--galaxy lensing signal provides a direct measurement of the total projected mass distribution around lens galaxies, including contributions from both the stellar and dark matter components. By modelling these components simultaneously, we can constrain the relative normalization of the stellar mass through the stellar mismatch parameter, $\alpha$.

Our composite model consists of an NFW halo profile describing the extended dark matter distribution and a scaled stacked stellar mass profile derived directly from the observed light distribution of the lens galaxies. In this framework, the shape of the stellar mass profile is fixed by the observed galaxy light profile, while the parameter $\alpha$ controls its overall normalization. This allows us to quantify deviations between the stellar mass inferred from the adopted stellar population synthesis model and the true stellar mass inferred via lensing measurements, which is a complementary probe to combined strong lensing and dynamical analyses \citep[e.g.,][]{2010_Treu}.

\cref{fig:alpha_constraints}, the blue points present the constraints on $\alpha$ as a function of stellar mass bins in the range $9.5 \leq \log(M_*/h^{-2} M_\odot) \leq 11.5$. The strongest constraints are obtained for intermediate stellar masses, where the combination of a sufficiently large number of lenses and a stronger lensing signal results in a relative precision of approximately 20 per cent per bin. Towards the low- and high-mass ends of the sample, the constraints become progressively weaker. This behaviour is driven by the reduced signal-to-noise ratio, as shown in \cref{fig:snr_iso_cen} for these stellar mass regimes, arising from the smaller number of galaxies and the lower lensing signal amplitude.

To evaluate the physical significance of these measurements, we compare the inferred constraints with the expected shifts based on various assumptions regarding the stellar IMF. We use the Chabrier IMF as our reference model for stellar population synthesis estimates and present the expected offsets corresponding to the Salpeter and Kroupa IMFs as benchmark values. These IMF selections represent commonly used alternatives that encompass the range of stellar mass-to-light ratios derived from different stellar population models.

Within the stellar mass range $10.0 \leq \log(M_*/ h^{-2} M_\odot) \leq 11.0$, the precision of our lensing measurements is sufficient to distinguish between these IMF prescriptions. At lower and higher stellar masses, the statistical uncertainties increase, and the measurements no longer provide strong constraints on $\alpha$. The $\alpha$ constraint obtained by combining the full stellar mass range provides a precision of approximately 8.5 per cent on the stellar mismatch parameter.

We also investigate whether any redshift evolution of the stellar mismatch parameter can be constrained. To do so, we select galaxies in the stellar mass range $10.5 \leq \log(M_*/h^{-2} M_\odot) \leq 11.0$, where the lensing constraints are strongest, and divide the sample into three redshift bins: $0.0 < z \leq 0.1$, $0.1 < z \leq 0.2$, and $0.2 < z \leq 0.3$. The orange points in \cref{fig:alpha_constraints} denote the inferred constraints on $\alpha$. We find the expected behaviour of improved constraints as a function of redshift due to the larger number of lenses in the high-redshift bins, and by restricting our analysis to the high signal-to-noise regime. The combined measurements achieve a precision of approximately 8.6 per cent, providing sufficient sensitivity to investigate possible redshift evolution and distinguish between standard IMF prescriptions.

Finally, we test the robustness of the inferred constraints against uncertainties in the structural properties of the lens galaxies. We introduce an additional intrinsic scatter of 0.3 dex into the galaxy size--stellar mass relation and quantify its impact on the recovered stellar mismatch parameter. This test allows us to assess potential systematic shifts and identify any biases associated with the assumed fixed size-stellar mass relation. We found similar constraints as for our fiducial sample and described in \cref{appex:alpha}. 

We compare the constraining power of our measurements of the stellar mismatch parameter with that of previous studies. For massive galaxies ($\log(M_*/h^{-2} M_\odot) \gtrsim 10.5$), our constraints are precise enough to distinguish between Chabrier and Salpeter IMFs, which differ by $\approx 0.25$\,dex in $\log\alpha$. Joint strong-lensing and dynamical analyses of massive early-type galaxies constrain the sample-averaged mismatch parameter to $\approx 0.04$--$0.05$\,dex \citep[e.g.][]{2010_Treu, 2010_Auger}, while \citet{2015_Smith_russell} reach a precision of $\approx 12\%$ ($\approx 0.05$\,dex) for three nearby lenses, assuming old stellar populations. Spectroscopic studies based on gravity-sensitive absorption features can achieve a statistical precision of $\approx 7\%$, but systematic uncertainties in stellar population models can inflate the total uncertainty on the inferred mass-to-light ratios to $\approx 50\%$ \citep[e.g.][]{2012_Conroy_b, 2013_LaBarbera}. Combining statistical strong lensing with a weak-lensing prior on halo mass, \citet{2019_Sonnenfeld} constrain the average $\log\alpha$ to $\pm 0.11$\,dex. In comparison, our weak lensing-only constraints reach an overall precision of $\approx 8.5\%$ ($\sigma_{\log\alpha} \approx 0.04$\,dex), comparable to that of strong-lensing and dynamical studies and a factor of $\sim 3$ tighter than \citet{2019_Sonnenfeld}. Moreover, our measurements span a broad range in stellar mass and multiple redshift bins, and do not rely on dynamical modelling or gravity-sensitive spectral features, although they still depend on the reference stellar population model and the assumed dark-matter halo profile.

In our galaxy--galaxy lensing analysis, we model the dark matter contribution with the standard NFW profile rather than a generalised NFW profile \citep{2001_Wyithe} with a free inner slope. We find that, within our scale cuts, the lensing signal is largely insensitive to the inner slope. Moreover, when the inner slope is left free, it is degenerate with the stellar mismatch parameter, $\alpha$, similar to the degeneracy reported in strong-lensing and stellar-kinematics studies \citep[e.g.][]{2010_Treu, 2025_Sonnenfeld}. This degeneracy is substantial: in a sensitivity test, changing the inner slope by 20 per cent shifts $\alpha$ by 25 per cent. We also test the approach described in \cref{subsec:satellites_galaxies} for including the contribution of satellite galaxies, and find results that are marginally consistent with our fiducial analysis. Further details are given in \cref{appex:alpha}.

\section{Summary}\label{sec:conclusions}

We demonstrated the potential of \textit{Euclid} to measure the small-scale galaxy-galaxy lensing signal to constrain stellar masses and probe the stellar initial mass function (IMF) of galaxies. To do so, created mock catalogues that mimic the DESI-like BGS, which overlaps with approximately $9000\,\mathrm{deg}^2$ of the full \textit{Euclid} imaging survey. Our analysis involved simultaneously modelling the stellar mass contribution and the surrounding dark matter halo over projected radial scales of $10$ to $100\,h^{-1}\,\mathrm{kpc}$ to constrain the stellar mismatch parameter, $\alpha$. 

To assess the impact of contamination by the light from the foreground lenses, we created simulated images. These were used to examine the biases in the lensing signal, and the photometric measurements of the sources. Finally, we performed a standard galaxy-galaxy lensing analysis on the resulting shape catalogue to measure the excess surface mass density, $\Delta \Sigma$. Our main findings can be summarised as follows:

\begin{itemize}
    
    \item \textbf{Impact of foreground lens light:} We demonstrated that foreground lens light causes significant scale-dependent systematic effects. This light reduces the detection of background sources and leads to negative additive biases in the measured weak-lensing signal at small projected separations. Consequently, these findings motivated our choice to exclude the inner region within four times the median half-light radius of the lenses in each stellar mass bin.
    \item \textbf{Biases in source photometry:} We demonstrated that the extended surface brightness of the foreground lens contaminated the photometry of background sources, resulting in magnitude offsets that led to systematic biases in photometric redshift estimates.

   \item \textbf{Mitigation through lens-light subtraction:} We modeled the foreground lens light using \textsc{Galfit}, which reduced these image-level systematic effects by about an order of magnitude, significantly improving the accuracy of the recovered weak-lensing signal, the detection efficiency, and the photometry of the sources.
    
    \item \textbf{Expected signal-to-noise ratio:} Our mock measurements provided high signal-to-noise ratios across the entire range of stellar masses considered, reaching approximately 150 for intermediate-mass central galaxies. This demonstrates that small-scale galaxy-galaxy lensing with \textit{Euclid} will provide the statistical precision needed to constrain stellar mass distributions on galactic scales.

    \item \textbf{Constraints on the stellar mismatch parameter:} We combined data across the entire stellar mass range of \(9.5 \leq \log(M_*/h^{-2} M_\odot) \leq 11.5\) to constrain the stellar mismatch parameter with an overall precision of approximately 8.5 percent. Additionally, we focused on massive central galaxies, specifically within the range \(10.5 \leq \log(M_*/h^{-2} M_\odot) \leq 11.0\), and by utilizing the redshift dependence of the signal, we achieved constraints at roughly the 8.6 percent level. These measurements are precise enough to distinguish between commonly used IMF prescriptions, such as Chabrier, Kroupa, and Salpeter IMFs. Furthermore, they facilitate the exploration of potential IMF redshift evolution.
    
\end{itemize}
To fully exploit small-scale galaxy--galaxy lensing in the \textit{Euclid} era, the theoretical framework should be extended to include baryonic gas and explicit models for satellite galaxies. Applying this approach to satellite galaxies will provide a complementary test of the flexibility of stellar population synthesis (SPS) models for galaxies residing in overdense environments.

In this work, we model the dark matter halos using an NFW profile, a standard choice for galaxy--galaxy lensing analyses at these scales \citep[e.g.][]{2006_Mandelbaum}. Since different IMF assumptions can change stellar masses by up to a factor of two, this simple parametrisation avoids introducing additional degeneracies with $\alpha$ and thus provides a convenient framework for our analysis. However, the inner slope of the dark matter profile is covariant with the stellar mismatch parameter; this degeneracy can be broken by combining weak lensing with probes that are more sensitive to the inner regions, such as strong lensing and stellar dynamics \citep[e.g.][]{2013_Newman, 2018_Sonnenfeld}. A more detailed understanding of the IMF will therefore require future work to incorporate more flexible halo models, while probing smaller scales to constrain the inner slope and address the core--cusp problem will require a thorough understanding of the relevant systematic effects.

Our methodology is also readily applicable to photometric lens samples, such as the KiDS Bright sample \citep{2021_Bilicki}. Applying this framework to such datasets will require robust joint modelling of photometric redshift and stellar mass uncertainties, including their covariance, to obtain accurate constraints on the stellar mismatch parameter.

Realizing the full scientific potential of these measurements will also require stringent control of observational systematics at small projected separations \citep{2018_Mandelbaum}. Future image simulations should therefore incorporate realistic source blending, accurate lens light profiles, and survey-specific PSFs, together with refined isolation criteria to mitigate environmental contamination \citep{2005_Mandelbaum}. Controlling these effects will establish small-scale galaxy--galaxy lensing as a powerful tool for probing the inner stellar and dark matter distributions of galaxies in the era of \textit{Euclid} \citep{2025_Mellier} and the \textit{Roman Space Telescope} \citep{2013_Spergel}.

\begin{acknowledgements}
DR thanks Preetish K. Mishra and Surhud More for the discussions related to \textsc{Galfit} and other aspects of this work. DR and HHo acknowledge funding from the European Research Council (ERC) under the European Union’s Horizon 2020 research and innovation program (Grant agreement No. 101053992). This work is financed in part by the Dutch Research Council (NWO) grant number OCENW.XL21.XL21.025.
AS acknowledges support from the National Key R\&D Program of China (No. 2023YFA1607800, 2023YFA1607802), and from the Yangyang Development Fund. This work has made use of CosmoHub, developed by PIC (maintained by IFAE and CIEMAT) in collaboration with ICE-CSIC. It received funding from the Spanish government (grant EQC2021-007479-P funded by MCIN/AEI/10.13039/501100011033), the EU NextGeneration/PRTR (PRTR-C17.I1), and the Generalitat de Catalunya. During the preparation of this work, AI tools were used to improve readability and restructure complex sentences.
\end{acknowledgements}

\bibliographystyle{aa}
\bibliography{apssamp}

\begin{appendix}

\section{Impact of blending on the source sample}
\label{app:blending}

\begin{figure}
    \centering
    \includegraphics[width=\linewidth]{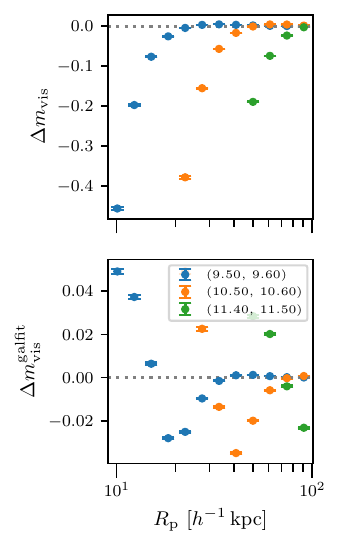}
    \caption{\textit{Impact of foreground lens light on source photometry.} The top panel shows the residual bias in the measured source magnitudes before lens-light subtraction, while the bottom panel shows the corresponding residual after modelling and subtracting the foreground lens light using \textsc{Galfit}. The source magnitudes are measured using the \texttt{MAG\_AUTO} estimator in \textsc{SExtractor}. The colour scheme and uncertainty estimates are identical to those adopted in \cref{fig:esd_ll_sys}.}
    \label{fig:stk_srcmag_ll_sys}
\end{figure}
Foreground lens light affects not only the measured source shapes but also the subsequent flux measurements and the probability of detection. Here, we explore the impact of the latter two, which may introduce additional scale-dependent systematic uncertainties. We quantify both effects using the same image simulations and lens-light subtraction procedure as used for the $\Delta\Sigma$ measurements.

\Cref{fig:stk_srcmag_ll_sys} illustrates the effect of lens light on the photometry of background sources. We quantify this effect through the bias in the source apparent magnitudes measured with the \texttt{MAG\_AUTO} estimator in \textsc{SExtractor} \citep{1996_Bertin}. Without lens-light subtraction (top panel), blending with the bright central galaxy introduces a systematic bias that artificially brightens nearby sources. These magnitude offsets follow the same stellar-mass dependence as found for $\Delta\Sigma$. In contrast, the bottom panel shows that modelling and subtracting the foreground light with \textsc{Galfit} reduces these photometric biases across all scales.

Such photometric biases matter because they propagate into the inferred source redshift distributions, introducing additional scale-dependent photo-$z$ uncertainties in galaxy-galaxy lensing analyses \citep{2012_Nakajima}. By removing them, lens-light subtraction should therefore also mitigate this source of photo-$z$ uncertainty, consistent with \citet{2023_Langeroodi}, who demonstrated that modelling and removing foreground lens light improves the recovery of source photometric properties.

Finally, we examine the detection efficiency of sources around the lenses. Without lens-light subtraction, we find a percent-level loss in detection efficiency, which decreases to the sub-percent level after lens-light modelling and subtraction. The scale-dependent detection bias is thus mitigated, and the recovered source population closely matches that obtained in the absence of foreground lens contamination.

\section{Signal-to-noise ratio}\label{appex:snr}

\begin{figure}
    \centering
    \includegraphics[width=\linewidth]{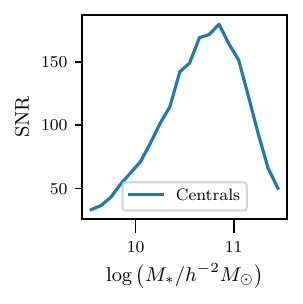}
    \caption{\textit{Signal-to-noise for the galaxy-galaxy lensing:} The blue solid line in the figure above represents the signal-to-noise ratio of weak lensing measurements as a function of stellar mass for our fiducial analysis of central galaxies.}
    \label{fig:snr_iso_cen}
\end{figure}

The statistical power of our measurements is significantly influenced by the lens stellar mass, as illustrated by the expected signal-to-noise ratio (SNR) for our fiducial sample of central galaxies in \cref{fig:snr_iso_cen}. We present the expected SNR for a lens sample divided into 27 stellar mass bins within the range $9.5 \leq \log(M_*/h^{-2} M_\odot) \leq 11.5$, assuming a DESI-BGS like survey overlapping the complete \textit{Euclid} footprint. This variation in SNR reflects the combined effects of the intrinsic signal strength, which increases with stellar mass, and the statistical uncertainty, which depends on the number of lenses in each bin.

\section{Variation in constraint by sample choice}\label{appex:alpha}

\begin{figure}
    \centering
    \includegraphics[width=\linewidth]{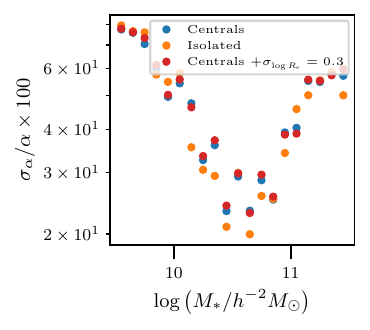}
    \caption{\textit{Effect of different sample choices:} The figure illustrates the constraints on the mismatch parameter $\alpha$ for various mock lens samples. The blue, orange, and red data points represent three scenarios: only central galaxies, isolated galaxies, and the effect of scatter on the adopted size-stellar mass relation.}
    \label{fig:alpha_const_nfw}
\end{figure}

As shown in \cref{fig:alpha_const_nfw}, constraints on the mismatch parameter vary depending on the sample selection. This is particularly evident when comparing the full central sample to one that applies isolation criteria and accounts for satellite galaxies, or when introducing a 0.3 dex scatter in the size–stellar mass relation. The blue, orange, and red data points illustrate the precision of these constraints, which generally show good agreement. However, we observe slight deviations for the isolated galaxy sample. These differences arise from the initial inclusion of satellite galaxies, followed by the application of isolation criteria, which ultimately alters the total number of lenses per stellar mass bin.

\end{appendix}

\end{document}